\pdfoutput=1  
\documentclass[letterpaper]{article} 
\usepackage{aaai2027}   
\nocopyright            
\usepackage[hyphens]{url}  
\usepackage{graphicx} 
\def\UrlFont{\rm}  
\usepackage{natbib}  
\usepackage{caption} 
\usepackage{amsmath}
\usepackage{amssymb}
\usepackage{booktabs}
\usepackage{tikz}
\usepackage{pgfplots}
\usetikzlibrary{arrows.meta,positioning,shapes.geometric}
\pgfplotsset{compat=1.18}

\title{The Vocabulary Gap Is an Equity Gap: Register Mismatch in Retrieval Systems for Public-Benefits Access}

\author{
    Krish Sapru
}
\affiliations{
    Dartmouth College \\
    krish.sapru.th@dartmouth.edu
}

\begin{document}
\maketitle

\begin{abstract}

Retrieval-augmented question answering is increasingly used to help people
navigate public-benefits eligibility, yet the documents these systems retrieve
from are written in agency register while intended users often ask questions in
plain, informal, or non-native English. We show that this register mismatch can
turn a high-performing retrieval system into an inequitable one. We construct a
controlled benchmark of 51 publicly documented federal benefit-eligibility
rules and 25 information needs, each phrased in both agency register and plain
user register while keeping the gold passage fixed. Across BM25, TF-IDF, and a
term-graph retriever, formal-register evaluation is nearly perfect
(Recall@5 96--100\%), but plain-register retrieval collapses (Recall@5
36--44\%). For BM25, Recall@1 falls from 84\% to 16\% and Recall@5 from
100\% to 44\%, a 56-point equity gap on identical information needs. We trace
the mechanism to a measurable vocabulary gap: formal queries share 0.63 of
their content terms with the gold passage, while plain queries share only 0.11,
a $5.9\times$ reduction. A deliberately simple, auditable plain-to-formal
lexicon bridge recovers much of the failure, lifting plain-query BM25 Recall@5
from 44\% to 80\%. The contribution is not a new retriever; it is an evaluation
protocol, benchmark, mechanistic diagnosis, and transparent mitigation for a
high-stakes social-impact failure mode that standard retrieval evaluation
hides.
\end{abstract}

\begin{links}
    {\def\UrlBreaks{}\def\UrlBigBreaks{}\def\UrlNoBreaks{}\def\UrlFont{\rm\scriptsize}%
     \renewcommand{\link}[2]{\par\textbf{#1}\newline\mbox{\url{#2}}}%
     \link{Code and benchmark}{https://github.com/ksapru/vocabulary-gap-code}}
\end{links}

\section{Introduction}

Access to public benefits is gated by information. A person is entitled to food
assistance, medical coverage, rental help, heating assistance, disability
income, or refundable tax credits only if they can discover the program,
determine whether they qualify, and learn how to apply. This information work
has long been shaped by administrative burden: learning costs, compliance
costs, and psychological costs that make formally available benefits
practically unavailable to many eligible people~\cite{herd2018}. The stakes are
large. The Legal Services Corporation reports that 92\% of the civil legal
problems of low-income Americans receive no or insufficient legal
help~\cite{lsc2022}. Benefits access is therefore not merely a retrieval
problem. It is an access-to-justice problem mediated increasingly by software.

Retrieval-augmented generation (RAG) systems promise to make this information
work scalable by retrieving relevant passages from statutes, eligibility
manuals, agency guidance, and nonprofit knowledge bases before generating an
answer~\cite{lewis2020}. But those source documents are written in
\emph{agency register}: the precise, controlled, statute-derived language of
program rules. The intended user often asks in another register entirely:
``can I still get food stamps if I work part time,'' ``my heat is about to get
shut off,'' or ``do I get money back at tax time if I barely made anything.'' A
system evaluated only on agency-like queries can appear safe and accurate while
failing the very users it is meant to serve.

This paper isolates that failure. We hold the information need and gold passage
fixed, vary only the register of the query, and measure the resulting equity
gap. The central result is stark: on identical information needs, three
standard retrievers perform well in the register of agencies and poorly in the
register of benefit-seekers. This is not a subtle ranking degradation. In the
BM25 system, formal Recall@5 is 100\%; plain Recall@5 is 44\%. Fourteen of 25
information needs that are answerable in agency register are lost when phrased
plainly.

We make four contributions.
\begin{itemize}
\item A paired-register evaluation protocol for benefits-access retrieval that
holds the information need fixed while varying only the query register.
\item A compact benchmark of 51 public benefit-eligibility rules and 25 paired
information needs spanning SNAP, Medicaid and ACA coverage, SSI/SSDI, TANF,
LIHEAP, housing assistance, WIC, school meals, and refundable tax
credits~\cite{usda2026,cms2023,hud2026,irs2026}.
\item Evidence of a large, method-agnostic equity gap across lexical,
vector-space, and graph-based retrieval.
\item A transparent mitigation: a small plain-to-formal lexicon bridge that
recovers most of the lost plain-register Recall@5 without an opaque model or
user data.
\end{itemize}

\section{Related Work}

\textbf{Retrieval and RAG.} Sparse lexical retrieval remains a strong baseline
in open-domain QA and RAG systems~\cite{robertson2009}. Dense passage retrievers
learn semantic representations and can outperform sparse baselines on many QA
benchmarks~\cite{karpukhin2020}, while RAG systems combine retrieved
non-parametric memory with sequence generation~\cite{lewis2020}. Graph-based
RAG methods construct graph indexes or entity summaries to improve
corpus-level sensemaking~\cite{edge2024}. Our finding is orthogonal to these
algorithmic advances: when both documents and evaluation queries use agency
language, retrieval quality is overestimated for users who do not use that
language.

\textbf{Vocabulary mismatch.} The vocabulary problem in human-system
communication is longstanding: different people choose different words for the
same object or task~\cite{furnas1987}. IR has addressed this through query
expansion, statistical translation, relevance feedback, and learned
representation methods~\cite{berger1999,karpukhin2020}. Our contribution is not
to rediscover mismatch, but to show that in benefits access the mismatch aligns
with social vulnerability. Agency vocabulary is not just a synonym set; it is
a barrier that distinguishes institutionally fluent users from users with less
legal, bureaucratic, or English-language fluency.

\textbf{Administrative burden and access to justice.} Administrative burden
research shows that policy implementation often shifts learning and compliance
costs onto citizens~\cite{herd2018}. Automated eligibility and public-service
systems can reproduce or amplify these burdens when design choices are not
accountable to affected communities~\cite{eubanks2018}. Legal-aid and
access-to-justice work similarly emphasizes that many low-income people do not
recognize their problems as legal or benefit-related problems at
all~\cite{lsc2022}. We connect this literature to retrieval evaluation: a
benefits assistant that understands only agency register lowers burden for the
already fluent while preserving it for the least resourced.

\textbf{Evaluating AI for social impact.} Work on AI for social impact
emphasizes problem formulation, data collection, field evaluation, social
significance, and engagement with literature beyond computer science. This
paper is written to that standard: it contributes a diagnostic benchmark and a
release-blocking metric for systems deployed in social-impact settings, rather
than optimizing a new model in isolation.

\section{Problem Formulation}

Let $C=\{p_1,\ldots,p_N\}$ be a corpus of benefit-rule passages written in
agency register. An information need $n$ has a gold passage set $g(n)\subseteq
C$. A user realizes $n$ as a query $q_r$ in register $r\in\{\text{formal},
\text{plain}\}$. A retriever returns an ordered list $R(q_r)$. For a metric
$M$, standard evaluation estimates $M_{\text{formal}}$ using queries that
resemble the corpus. The equity-relevant quantity is instead
\begin{equation}
\Delta_M = M_{\text{formal}} - M_{\text{plain}},
\end{equation}
where both metrics are computed over the same needs and gold passages. A system
can have excellent $M_{\text{formal}}$ and a large $\Delta_M$: it works in the
lab and fails in deployment.

\section{Corpus and Benchmark}

\textbf{Corpus.} We assemble 51 passages, each stating a single publicly
documented federal benefit-eligibility rule. The corpus spans SNAP, Medicaid
and ACA marketplace coverage, SSI/SSDI, TANF, LIHEAP, Housing Choice Vouchers
and public housing, WIC and school meals, EITC and Child Tax Credit, plus
cross-cutting procedural rules such as recertification, expedited processing,
fair hearings, verification, and mixed-status households. Public guidance from
federal agencies and benefit administrators uses the controlled terms that
motivated the study: for example, SNAP guidance uses ``Supplemental Nutrition
Assistance Program'' and eligibility thresholds~\cite{usda2026}; Medicaid
eligibility guidance uses MAGI and federal-poverty-level
terminology~\cite{cms2023,healthcare2026}; Housing Choice Voucher guidance uses
program-specific voucher language~\cite{hud2026}; EITC guidance uses
qualifying-child, earned-income, and valid Social Security number
terminology~\cite{irs2026}.

\textbf{Paired queries.} We author 25 information needs. Each need is phrased
twice: once in formal agency register and once in everyday plain register,
including informal program names, incomplete grammar, and terms a
benefit-seeker might actually type. For example, a formal query may ask about
LIHEAP crisis assistance, while its paired plain query asks whether help is
available when ``my heat is about to get shut off.'' Both queries map to the
same gold passage(s). The benchmark therefore contains 50 queries over 25
controlled needs.

\textbf{Why author queries rather than scrape logs?} Live benefits-search logs
would be more realistic, but they may contain private, legally sensitive, or
immigration-relevant information. Using authored pairs is a deliberate ethical
choice for a first benchmark. It allows causal isolation of register while
avoiding the collection of vulnerable users' data. We treat this as a
lower-bound protocol, not a replacement for future IRB-governed,
community-consented query collection.

\section{Methods}

We evaluate three retrievers, all standard and unmodified.

\textbf{BM25.} Okapi BM25 over tokenized passages is a strong lexical baseline
and remains competitive in many retrieval settings~\cite{robertson2009}.

\textbf{TF-IDF vector similarity.} The TF-IDF retriever ranks passages by
cosine similarity between query and passage vectors, representing a classic
vector-space lexical model.

\textbf{Term-graph retrieval.} The graph retriever constructs a term
co-occurrence graph with passage and term nodes. Query terms seed personalized
PageRank, and retrieval reads off passage node mass. This is not a full
LLM-based GraphRAG system; it is a deliberately transparent graph baseline
designed to test whether corpus structure can rescue vocabulary mismatch when
no learned encoder is introduced.

\textbf{Metrics.} We report Recall@1, Recall@3, Recall@5, and MRR@10 per
register. The equity gap is formal minus plain. Because $N=25$, recall
percentages correspond directly to counts in increments of four percentage
points. For the main BM25 Recall@5 gap, all 25 formal queries retrieve a gold
passage in the top five, while only 11 of 25 plain queries do. This means 14
paired needs are lost under register shift. The asymmetry is near-total: of the
14 discordant pairs (one register succeeds, the other fails), all 14 favor the
formal register and none favor the plain register (exact two-sided sign test
$p\approx1.2\times10^{-4}$). Because the benchmark is authored and paired by
construction, we read this as evidence that the register effect is systematic
rather than incidental, not as an estimate of population-level prevalence.

\begin{figure*}[t]
\centering
\begin{tikzpicture}[font=\small, node distance=5mm and 7mm,
box/.style={draw, rounded corners, align=center, minimum width=2.5cm, minimum height=0.9cm, fill=gray!8},
arr/.style={-{Latex[length=2mm]}, thick}]
\node[box] (need) {Information need\\$n$};
\node[box, right=of need] (formal) {Formal query\\agency register};
\node[box, below=of formal] (plain) {Plain query\\user register};
\node[box, right=of formal, yshift=-0.45cm] (retr) {Same retriever\\same corpus};
\node[box, right=of retr] (rank) {Ranked passages};
\node[box, right=of rank] (metrics) {Recall@$k$, MRR\\and gap $\Delta$};
\node[box, below=of retr, yshift=-0.15cm] (bridge) {Auditable bridge\\folk $\rightarrow$ agency terms\\query expansion};
\draw[arr] (need) -- (formal);
\draw[arr] (need.east) |- (plain.west);
\draw[arr] (formal) -- (retr);
\draw[arr] (plain) -- (retr);
\draw[arr] (retr) -- (rank);
\draw[arr] (rank) -- (metrics);
\draw[arr] (plain) -- (bridge);
\draw[arr] (bridge) -- (retr);
\node[below=6mm of need, align=left, font=\footnotesize] {Hold gold passage\\ fixed; vary only register.};
\end{tikzpicture}
\caption{Paired-register evaluation protocol. Each information need is realized
as both a formal agency-register query and a plain user-register query. Because
the corpus, retriever, and gold passage are fixed, differences in retrieval
quality estimate the register-induced equity gap rather than topical
difficulty.}
\label{fig:protocol}
\end{figure*}
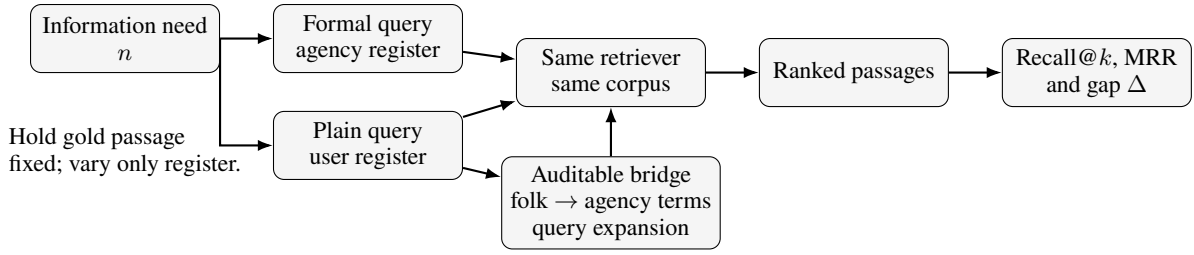

\section{Results}

\subsection{A large, consistent equity gap}

Table~\ref{tab:main} reports retrieval performance for all three systems. The
pattern is consistent across methods. Formal-register performance is high:
Recall@5 is 96--100\%, and MRR is at least 0.817. Plain-register performance on
the same needs collapses: Recall@5 falls to 36--44\%, and Recall@1 falls to
12--24\%. The graph retriever provides no rescue and is weakest on plain
queries, which is expected if graph structure is built from the same agency
terms that the user does not supply.

\begin{table}[t]
\centering
\small
\begin{tabular}{llrrrr}
\toprule
Retriever & Register & R@1 & R@3 & R@5 & MRR \\
\midrule
BM25 & formal & 84.0 & 100.0 & 100.0 & 0.913 \\
 & plain & 16.0 & 32.0 & 44.0 & 0.295 \\
 & \textbf{GAP} & \textbf{68.0} & \textbf{68.0} & \textbf{56.0} & \textbf{0.618} \\
\midrule
TF-IDF & formal & 88.0 & 96.0 & 100.0 & 0.930 \\
 & plain & 24.0 & 32.0 & 44.0 & 0.334 \\
 & \textbf{GAP} & \textbf{64.0} & \textbf{64.0} & \textbf{56.0} & \textbf{0.596} \\
\midrule
Graph & formal & 72.0 & 88.0 & 96.0 & 0.817 \\
 & plain & 12.0 & 20.0 & 36.0 & 0.243 \\
 & \textbf{GAP} & \textbf{60.0} & \textbf{68.0} & \textbf{60.0} & \textbf{0.574} \\
\bottomrule
\end{tabular}
\caption{Retrieval on formal vs. plain phrasings of identical
information needs. GAP = formal $-$ plain; higher gap means larger equity
harm.}
\label{tab:main}
\end{table}

\subsection{Which needs are lost}

For BM25, 14 of 25 information needs are answerable in formal register but lost
in plain register. The failures are not edge cases. They include common,
high-stakes questions: whether income is low enough for food assistance,
whether a college student qualifies, how quickly emergency help arrives,
whether a green-card holder must wait, whether working parents can get
child-care help, and whether low earners receive money back at tax time. These
are exactly the questions where a silent retrieval failure can cause a family
not to apply for a benefit it may be entitled to.

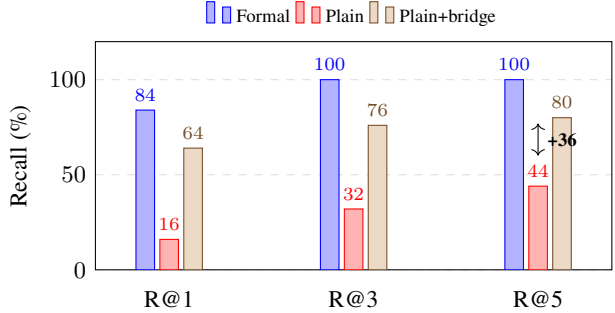
\begin{figure}[t]
\centering
\begin{tikzpicture}
\begin{axis}[
 ybar, width=\columnwidth, height=4.6cm, bar width=7pt,
 ymin=0,ymax=120,ylabel={Recall (\%)},
 symbolic x coords={R@1,R@3,R@5},xtick=data,
 enlarge x limits=0.20,ymajorgrids=true,grid style={dashed,gray!25},
 tick style={draw=none},legend style={at={(0.5,1.02)},anchor=south,
 legend columns=3,draw=none,font=\scriptsize},
 tick label style={font=\small},label style={font=\small},
 nodes near coords,nodes near coords style={font=\scriptsize}]
\addplot coordinates {(R@1,84) (R@3,100) (R@5,100)};
\addplot coordinates {(R@1,16) (R@3,32) (R@5,44)};
\addplot coordinates {(R@1,64) (R@3,76) (R@5,80)};
\legend{Formal,Plain,Plain+bridge}
\draw[<->,thin] (axis cs:R@5,60) -- node[right,font=\scriptsize\bfseries] {+36} (axis cs:R@5,77);
\end{axis}
\end{tikzpicture}
\caption{BM25 retrieval under formal, plain, and bridged plain queries.
Register shift causes a 56-point Recall@5 gap; a transparent lexicon bridge
recovers 36 points, closing roughly 64\% of the observed Recall@5 gap.}
\label{fig:recovery}
\end{figure}

\subsection{Mechanism: the vocabulary gap}

The cause is lexical and measurable. Formal queries share a mean 0.63 of their
content terms with the gold passage, while plain queries share only 0.11. The
result is a $5.9\times$ reduction in query-gold term overlap. The words users
type---``food stamps,'' ``welfare,'' ``daycare,'' ``green card,'' ``heat shut
off''---are often absent from the agency vocabulary, which says
``Supplemental Nutrition Assistance Program,'' ``Temporary Assistance for
Needy Families,'' ``Child Care and Development Fund,'' ``lawfully present
noncitizen,'' and ``Low Income Home Energy Assistance Program.''

Lexical retrievers cannot match terms that never occur. The term-graph retriever
inherits the same blind spot because its graph is induced from the same agency
corpus. This explains why the gap is method-agnostic across the evaluated
systems.

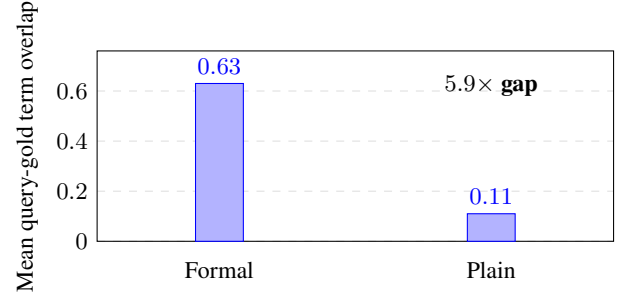
\begin{figure}[t]
\centering
\begin{tikzpicture}
\begin{axis}[
 ybar, width=\columnwidth, height=4.1cm, bar width=18pt,
 ymin=0,ymax=0.76,ylabel={Mean query-gold term overlap},
 symbolic x coords={Formal,Plain},xtick=data,
 enlarge x limits=0.45,ymajorgrids=true,grid style={dashed,gray!25},
 tick style={draw=none},nodes near coords,
 nodes near coords style={font=\small},
 tick label style={font=\small},label style={font=\small}]
\addplot coordinates {(Formal,0.63) (Plain,0.11)};
\node[font=\small\bfseries] at (axis cs:Plain,0.62) {$5.9\times$ gap};
\end{axis}
\end{tikzpicture}
\caption{Mechanism of failure. Plain-register queries share far fewer terms
with the correct agency passage than formal-register queries do. The equity gap
is therefore not mysterious model behavior; it is a measurable vocabulary
barrier.}
\label{fig:overlap}
\end{figure}

\subsection{A cheap, auditable remedy}

We test a deliberately simple mitigation: a hand-curated plain-to-formal
lexicon bridge applied as query expansion. Examples include ``food stamps''
$\rightarrow$ ``supplemental nutrition assistance program,'' ``welfare''
$\rightarrow$ ``temporary assistance for needy families,'' ``daycare''
$\rightarrow$ ``child care and development fund,'' and ``heat shut off''
$\rightarrow$ ``low income home energy assistance.'' The bridge is not a
black-box rewrite. It is a small, inspectable artifact that a legal-aid
organization or benefits agency could maintain, audit for bias, and update as
local language changes.

\begin{table}[t]
\centering
\small
\begin{tabular}{lrrrr}
\toprule
Plain-query BM25 & R@1 & R@3 & R@5 & MRR \\
\midrule
Baseline & 16.0 & 32.0 & 44.0 & 0.295 \\
+ bridge & 64.0 & 76.0 & 80.0 & 0.722 \\
$\Delta$ & +48.0 & +44.0 & +36.0 & +0.427 \\
\bottomrule
\end{tabular}
\caption{Plain-language BM25 retrieval with and without the lexicon
bridge.}
\label{tab:bridge}
\end{table}

The bridge lifts plain-query Recall@5 from 44\% to 80\% and MRR from 0.295 to
0.722. Put differently, it recovers nine additional top-five successes out of
25 needs. The bridge and the plain-register queries were authored by the same
process, so this number is best read as an upper bound on what a hand-curated
lexicon recovers, not as an estimate of held-out performance. Its importance is
institutional rather than methodological: a transparent, low-cost, auditable
intervention can close much of the measured equity gap before an organization
reaches for more opaque neural rewriting.

\section{Discussion}

\textbf{Why standard evaluation hides the harm.} A retrieval benchmark whose
queries are written by system builders, lawyers, or agency experts is likely to
share vocabulary with the corpus. Reporting only formal-register metrics
therefore certifies a system as accurate under the language of institutions,
not under the language of users. In benefits access, that is an equity problem
because the users least likely to know agency terms are also the users least
able to absorb a silent false negative.

\textbf{Release-blocking metric.} We recommend that benefits-access RAG systems
report the register gap $\Delta_M$ alongside aggregate retrieval quality. A
system should not be shipped only because $M_{\text{formal}}$ is high. A large
$\Delta_M$ means that the system is conditionally accurate: it works for users
who already speak the institution's vocabulary. In deployment, the gap should
be treated as a release-blocking metric, analogous to an error-rate threshold
for a vulnerable subpopulation.

\textbf{Why transparent bridges matter.} Neural query rewriting may eventually
improve this task, but social-impact systems require contestability and
auditability. If ``food stamps'' is mapped to SNAP, a legal-aid lawyer can
inspect the mapping. If a model rewrites a query into the wrong program, the
failure is harder to audit and explain. A lexicon bridge is therefore not
merely a baseline. It is a governance-compatible design choice for high-stakes
public-service retrieval.

\textbf{Generalization beyond benefits.} The same pattern is likely wherever
institutions use controlled vocabularies and users use lived vocabularies:
immigration assistance, tax credits, housing rights, disability accommodations,
unemployment insurance, and student aid. The paired-register protocol can be
reused in any domain where the question is not only ``does retrieval work?''
but ``for whom does retrieval work?''

\section{Limitations and Ethics}

The benchmark is modest in scale and authored rather than collected from live
logs. This limits claims about population-level prevalence. However, the paired
design strengthens causal interpretation: each formal/plain pair expresses the
same need and maps to the same gold passage. The plain phrasings approximate,
but do not fully capture, the diversity of low-literacy, multilingual,
disability-related, and dialectal inputs. A deployed benchmark should be built
with community organizations under informed consent and privacy review.
The corpus encodes benefit-eligibility rules as publicly documented at the time
of writing; program rules change through legislation and regulation (for
example, recent statutory changes to SNAP work and non-citizen eligibility
provisions), so any deployed system must track the authoritative current rules.
Our contribution concerns \emph{retrieval} over a fixed corpus, not the currency
of benefit advice, and the register-gap findings are independent of which
specific rules the corpus encodes.

The three retrievers we evaluate are all lexical in the relevant sense: BM25
and TF-IDF match surface terms directly, and the term graph is induced from the
same agency corpus, so it inherits the same vocabulary. We therefore cannot
claim the gap persists under learned dense representations, which are designed
to bridge exactly this kind of surface mismatch. Establishing whether a dense
retriever narrows, preserves, or merely relocates the register gap is the most
important open question left by this study.

The lexicon bridge is illustrative. If maintained carelessly, it could encode
regional, racial, or linguistic blind spots. For that reason, we recommend that
bridges be versioned, publicly documented, evaluated by subgroup-relevant query
sets where ethically available, and co-designed with legal-aid or benefits
navigators. We also do not claim that retrieval success alone guarantees
correct benefit advice. Retrieved passages must still be used by
answer-generation systems with citation, uncertainty, jurisdictional caveats,
and referral to qualified assistance where appropriate.

\section{Conclusion}

Retrieval systems increasingly stand between low-income people and the benefits
they may be entitled to. We showed that standard evaluation in agency register
can make these systems appear reliable while they fail on the plain-language
queries intended users actually type. Across the lexical retrievers we
evaluate, the failure is large, mechanistically explained by a $5.9\times$
vocabulary-overlap gap, and partly remediable by cheap, auditable query
expansion. The core lesson for AI for social impact is procedural: measure the
register gap before deployment. In a domain where a missed passage can mean a
family does not learn it qualifies for food, medical care, rent support, or
emergency heating help, a retrieval system that works only for the fluent is
not a social-impact success.

\bibliography{references}

\end{document}